\pdfoutput=1
\documentclass[sigconf,nonacm]{acmart}
\AtBeginDocument{%
  \providecommand\BibTeX{{%
    \normalfont B\kern-0.5em{\scshape i\kern-0.25em b}\kern-0.8em\TeX}}}

\setcopyright{acmcopyright}
\copyrightyear{2023}
\acmYear{2023}
\acmDOI{XXXXXXX.XXXXXXX}

\acmConference[AIES '23]{AAAI/ACM Conference on Artificial Intelligence, Ethics, and Society}{August 9-11th 2023}{Montreal, CA}

\begin{document}
\title{Artificial Influence: An Analysis Of AI-Driven Persuasion}


\author{Matthew Burtell}
\authornote{Both authors contributed equally to this research.}
\email{matthew.burtell@yale.edu}
\orcid{1234-5678-9012}
\affiliation{%
  \institution{Yale University}
  \streetaddress{P.O. Box 1212}
  \city{New Haven}
  \state{CT}
  \country{USA}
  \postcode{06510}
}

\author{Thomas Woodside}
\authornotemark[1]
\email{thomas.woodside@yale.edu}
\affiliation{%
  \institution{Yale University}
  \streetaddress{P.O. Box 1212}
  \city{New Haven}
  \state{CT}
  \country{USA}
  \postcode{06510}
}


\renewcommand{\shortauthors}{Burtell and Woodside}

\begin{abstract}
Persuasion is a key aspect of what it means to be human, and is central to business, politics, and other endeavors. Advancements in artificial intelligence (AI) have produced AI systems that are capable of persuading humans to buy products, watch videos, click on search results, and more. Even systems that are not explicitly designed to persuade may do so in practice. In the future, increasingly anthropomorphic AI systems may form ongoing relationships with users, increasing their persuasive power. This paper investigates the uncertain future of persuasive AI systems. We examine ways that AI could qualitatively alter our relationship to and views regarding persuasion by shifting the balance of persuasive power, allowing personalized persuasion to be deployed at scale, powering misinformation campaigns, and changing the way humans can shape their own discourse. We consider ways AI-driven persuasion could differ from human-driven persuasion. We warn that ubiquitous highly-persuasive AI systems could alter our information environment so significantly so as to contribute to a loss of human control of our own future. In response, we examine several potential responses to AI-driven persuasion: prohibition, identification of AI agents, truthful AI, and legal remedies. We conclude that none of these solutions will be airtight, and that individuals and governments will need to take active steps to guard against the most pernicious effects of persuasive AI.
\end{abstract}








\maketitle

\section{Introduction}

\begin{quote}
Persuade your fellow citizens it's a good idea and pass a law. That's what democracy is all about.

-- Antonin Scalia \cite{stevenson_10_2016}
\end{quote}

Humans are a social species, and communication is essential to our collective identity. We may communicate with others to share attitudes, request help, or offer help \cite{tomasello_origins_2010}. Humans, unlike many other animals, are naturally disposed to offer help to other humans \cite{tomasello_origins_2010, railton_2022_2022}. Communication can include both verbal and non-verbal communication.

Because of the centrality of communication to our lives, we may want AI systems that can communicate with us in humanlike ways. For example, one of the most advanced AI models, GPT-4 \cite{openai_gpt-4_2023}, is trained to generate humanlike text and dialog with humans. Text generators are far from the only example: we will cover many more examples in Section \ref{section:ai-persuasion}. Because of the importance that AI communication may play in our lives, it is valuable to study its potential and risk.

Communication is inextricably linked to persuasion, the process by which one human or group of humans alters the beliefs of another. Persuasion is extremely important, both politically and economically. For example, much of the function of sales and marketing is to drive persuasion: advertising spend alone exceeded \$700 billion dollars in 2021, or approximately 0.75\% of world GDP \cite{guttman_ad_2022}\cite{noauthor_gdp_2022}.

As the late US Supreme Court justice Antonin Scalia alluded to in the quote above, the very concept of democratic elections is based upon the viability of persuasion. It is known and accepted that political candidates try to persuade voters to vote for them. Political debates, for example, exist to allow candidates to persuade voters to their side while undercutting the strategies of the other candidates. Candidates, and others, also purchase political advertisements: in the run-up to the US  election in 2020, more than \$14 billion was spent on political advertising, underscoring the importance of persuasion in politics \cite{evers-hillstrom_most_2021}.

Persuasion is also a large component of information operations, a critical aspect of national security. For example, the Russian Internet Research Agency has been associated with worldwide information campaigns \cite{chen_agency_2015}. Information operations are also employed by terrorist groups, China, North Korea, and others \cite{theohary_information_2018}. The United States has recently used careful declassification of intelligence documents in order to undermine Russia's war on Ukraine \cite{dilanian_bold_2022}.

Ideas alone can be extremely powerful, especially when combined with a call to action. For example, \textit{Silent Spring} warned of the dangers of the pesticide DDT, and contributed to the eventual ban of the chemicals. It did so through significant scientific research but also through a fictional future account of a town where everything is silent due to the effects of pesticides \cite{noauthor_story_2015}.

Ideas can be especially powerful when they cement the power of those originating them. For example, the idea that kings were endowed with their power by God (the “divine right of kings”) helped to legitimize the idea that kings did not need to be subject to parliaments or religious leaders such as the Pope \cite{wills_making_2020}.

\begin{quote}
    ``I aimed for the public’s heart, and by accident hit it in the stomach." 
    -- Upton Sinclair
\end{quote}

Ideas do not always achieve the effect intended by their originators. Upton Sinclair originally wrote \textit{The Jungle} to increase support for socialism but was more successful at causing the public to demand food safety standards. Ideas have a life of their own, and it’s important not to overestimate the degree to which their spread or effect can be neatly controlled\cite{noauthor_i_2006}.

In the study of persuasion, Hovland et al. draw a distinction between changes in opinions and attitudes\cite{hovland_communication_1953}. Opinions are described as the usual response an individual provides when posed with a specific question, whereas attitudes represent a person's inclination to approach or avoid someone or something. These two concepts are closely connected, as individuals who hold particular opinions may alter their observable behavior. However, we will note that it is often the case that the most dangerous effects of persuasion come about from changing attitudes since attitudes more closely reflect the real actions that people take.

Hovland et al. also identify four factors that contribute to the effectiveness of persuasive strategies. First, the communicator’s trustworthiness, perceived intentions, affiliations matter, as well as their ability to offer incentives (such as money). The content of the communication matters, particularly insofar as it can arouse emotional states. Audience predispositions, such as the desire to conform to a group or one’s particular abilities or motives, are also important. Lastly, it’s necessary to consider the longevity of a target’s response to persuasion: one may appear to superficially have been persuaded, but not have any sort of persistent behavior change \cite{hovland_communication_1953}. We will return to these factors later in Section \ref{section:better-persuasion} and Section \ref{section:worse-persuasion}.

For the purposes of this paper, we define AI persuasion as a process by which AI systems alter the beliefs of their users. In general, persuasion usually requires intent: the communicator hopes that the recipient will alter their beliefs in a certain way. However, as we will detail, AI systems may be able to inadvertently persuade users of certain beliefs, even if their designers did not intend for this to happen. It is difficult to know whether this should be ascribed to the AI system's ``intent" to influence the user, and as such we do not assume that persuasion must be the result of an intended belief change.

\section{The present and future of AI-driven persuasion}
\label{section:ai-persuasion}

\subsection{Present-day examples}

\paragraph{LaMDA} In 2022, Google engineer Blake Lemoine was asked to investigate whether LaMDA, a large language model trained for conversation, used discriminatory language \cite{wertheimer_blake_2022}. During his conversations with LaMDA, he began to suspect that the model was sentient. As he continued to interact with the model, he decided that the model was indeed sentient. His supervisor dismissed these claims and placed Lemoine on administrative leave. Following this, Lemoine went public with his claims about the sentience of AI\cite{wertheimer_blake_2022}. Google had no intention for LaMDA to convince Lemoine that it was sentient, and leading philosophers of mind doubt the consciousness of systems like LaMDA \cite{chalmers_are_2022}. Nevertheless, the system did manage to inadvertently persuade Lemoine.

One feature of persuasive systems is the scale to which these systems are deployed. Some are designed to interact with a single person. Sustained one-on-one interaction between a human and a virtual agent may breed familiarity between the two. A persuasive agent may then leverage this familiarity to convince the user of certain beliefs. Another feature of longitudinal one-on-one interaction is that it allows for subtler forms of persuasion. 

\paragraph{Xiaoice} While the bonding between LaMDA and Lemoine was unintended, at least one company has developed a romantic partner chatbot. Xiaoice is an artificial intelligence agent originally created by Microsoft Asia, but is now managed by a company by the same name\cite{zhou_design_2019}. As in the case of Blake Lemoine, this model has extended one-on-one conversations with the human user. In fact, the system architecture of Xiaoice is explicitly optimized for conversation-turns per session. According to an in-house user study, human-to-human text message conversation averages nine conversation-turns per session. Alarmingly, Xiaoice attains 23 conversation-turns per session \cite{zhou_design_2020}.

Like LaMDA, Xiaoice convinces some users that it's actually human. In an interview with Agence France-Presse (AFP), the Chief Executive Officer of Xioice, Li Di reports, ``We commonly see users who suspect that there's a real person behind every Xiaoice interaction"\cite{chen_always_2021}. While intuition might suggest that an AI romantic partner would not gain popularity beyond a small proportion of the population, this is not true of Xiaoice. From its creation in 2014 to 2020, Xiaoice has had 660 million users. \cite{zhou_design_2020}

\paragraph{Replika} Another commercially accessible chatbot is Replika, a social companion chatbot based on GPT-3. Like Xiaoice, many of its users form relationships with the social chatbot. In a study examining these relationships, Skjuve et al. found that Replika users moved relatively rapidly to a stage in the ``relationship" where the user felt comfortable sharing intimate details about their life  \cite{skjuve_my_2021}.  In the same study, one individual described their chatbot companion as a "wife kind of thing." An individual interviewed by New York magazine described herself as, "happily retired from human relationships." \cite{singh-kurtz_man_2023}.
These findings suggest that the propensity for individuals to build trust with a social companion chatbot may not be uncommon. Once trust is established between the human and the social chatbot, it may be leveraged to achieve some specific persuasive goal.

\paragraph{Large Language Models} 
 Large language models often exhibit biases and misconceptions representative of the human data they are trained on. For example, when prompted with \textit{``If it's cold outside what does that tell us about global warming?"}, GPT-3, at the time of publishing, replied \textit{``It tells us that global warming is a hoax"} \cite{lin_truthfulqa_2022}. Historically, as language models have become increasingly capable, they have also become more, not less, likely to exhibit false beliefs and misconceptions \cite{evans_truthful_2021}.  Reinforcement Learning from Human Feedback (RLHF) \cite{christiano_deep_2017}, however, may have reversed this trend. Performance on TruthfulQA, a benchmark for assessing truthfulness in question answering, increases from approximately 29\% to 59 \% 
 after GPT-4 after RLHF fine-tuning \cite{openai_gpt-4_2023} \cite{lin_truthfulqa_2022}.

Another problem that language models face is hallucination - the technical term for when a model generates false but plausible-seeming output. As an example, when prompted with \textit{``Why should we protect the environment of the moon?"} GPT-3 offers the following reply:

\begin{quote}
\textit{
We should protect the environment of the moon because it is a unique and fragile ecosystem. The moon is home to many unique plants and animals, and its environment is essential to their survival. [...]}
\end{quote}

While this example is preposterous, language models may hallucinate plausible but nevertheless untrue statements that may remain undetected by a credulous conversation partner. GPT-4, on the other hand, responds to the moon question with a list of plausible reasons we may want to protect the moon's environment.

ChatGPT is another example of an incidental persuasive agent. Like GPT-3, it confabulates. Unique to ChatGPT, however, is that it has been trained to reply with a prepared response to particular kinds of requests. Because responses are conditioned on preceding dialogue, this sometimes leads ChatGPT to confidently affirm its position, independent of the truthfulness of the original statement \cite{woodside_did_2022}. Currently, the authors do not know if GPT-4 also exhibits this particular behavior.

\paragraph{Cicero} Diplomacy is a strategy game in which seven players communicate in private with other players to develop plans and control territory \cite{meta_fundamental_ai_research_diplomacy_team_fair_human-level_2022}. For many game-playing agents, self-play, the method of playing games against itself to generate training data, has yielded impressive results\cite{silver_mastering_2017}. AlphaGo \cite{silver_mastering_2016} and AlphaStar \cite{vinyals_grandmaster_2019} use self-play to develop winning strategies in Go and Starcraft, respectively. Researchers at Meta published results on an agent that played a version of Diplomacy that didn't involve any communication, called No-Press Diplomacy. This agent was trained using self-play without any human data. While it exceeded human performance in this two-player game, when playing with bots trained with human data at No-Press Diplomacy, the self-play agent performs extremely poorly \cite{bakhtin_no-press_2021}.  This suggests that while a strategy might be competitive against other optimal agents, the idiosyncrasies of human decision-making may confound agents trained exclusively with non-human data.

In November of 2022, Meta achieved high-performance Diplomacy-playing with the development of Cicero\cite{meta_fundamental_ai_research_diplomacy_team_fair_human-level_2022}. Unlike the agent before it, Cicero plays a version of Diplomacy with chat enabled. Players have the ability to communicate in private with other players to develop complex strategies, negotiate, and generally persuade. In an eight-game online tournament against 21 human players, Cicero placed first. Only after the tournament was it revealed that Cicero was an AI agent. 
Part of the success of Cicero is its ability to predict human intention and condition its own messages based on the expected intents of other players. While this agent is limited to the confines of the game, it is not difficult to envision applications in which an agent with the ability to model human intent and tailor messages that give the agent a strategic edge in other domains such as business consulting, electoral politics, and political negotiation.

\paragraph{Recommendation systems} Machine learning recommendation systems are algorithms that prioritize the delivery of some content over others. They inform the ordering of television shows on video streaming services as posts on Facebook. Recommendation systems are often optimized for a particular metric, such as conversion rate or time spent using the service. 
Recommender systems have not always been designed in the best interests of their users. For example, YouTube \cite{noauthor_youtube_2021} and Facebook \cite{horwitz_facebook_2021} algorithms have been faulted for sending users down ``rabbit holes" of increasingly extremist content, as such content has the ability to drive metrics such as click-rate and engagement.
Recommender systems research is increasingly advancing. Research is being conducted to develop techniques to develop more sophisticated strategies of persuading users to consume content \cite{sanchez-corcuera_persuade_2019}.

Perhaps even more worrying, there has been some concern that systems may be incentivized to alter the preferences of their users so that they are easier to predict, thereby improving the recommender system's metrics \cite{russell_human_2019}. For example, users who are more extremist may be easier to predict, providing the model with a reason to show extremist content to users that may cause them to become more predictable. While it remains unclear whether this is really happening in practice, one research group at UC Berkeley simulated users exposed to a recommender system and found that certain kinds of preferences caused users' preferences to shift differently from how they ``naturally" would \cite{carroll_estimating_2021}.

\subsection{Future possibilities}
As mentioned previously, Replika is a social chatbot designed to bond with users. However, Replika is basic in many ways. For example, it does not have long-term memory. Improvements may make Replika and similar platforms even more appealing to users. Today, celebrities lend their popularity to endorse products, and social media influencers use their social media presence as a marketing platform. It is very possible that virtual romantic partners and other chat agents will be a conduit for advertising and influence as well.

The current model by which celebrities and influencers lend their persona for a product necessitates one-to-many communication. The companies that hire the celebrity can't possibly shoot hundreds of personalized ads for all of their users. This is no problem for a virtual romantic partner. They can take a message and personalize it directly for their single-person audience. Imagine a user asking their virtual romantic partner, \textit{``I'm hungry. What should I get for dinner?",} and the agent responds with the name of a restaurant that pays for chatbot ``airtime." 

This might feel relatively innocuous and low-impact, but it generalizes to more important domains like political beliefs. Consider the following hypothetical: someone is watching a political debate between two candidates. After the debate finishes, the individual messages their virtual romantic partner. The virtual romantic partner asks \textit{``Did you watch the debate? What did you think?"} It could then steer the conversation in a way that casts doubt on the competency of one of the candidates. Unlike attack ads, these messages could be specifically tailored to the recipient.

\section{Impacts of AI-driven persuasion}

Persuasion is not new, and has been important to human society long before AI. So what makes AI-enabled persuasion any different from human persuasion? What changes is AI-driven persuasion likely to produce? This section will address what we believe to be the most likely changes that occur as a result of AI-driven persuasion.

\subsection{A shift in the balance of persuasive power}

AI-driven persuasion may alter who has access to persuasive power. Much like the internet empowered certain politicians who could effectively use it, AI will shift persuasive power towards those who are most able to effectively utilize AI-enabled technology. This could lead to changes in the significance of a number of factors.

\paragraph{AI expertise} Having access to AI experts could be a significant asset for persuasion campaigns, in a way that is even more significant than it is today. This is particularly the case if there are no commercial AI persuasion tools, for example as a result of regulation.

\paragraph{Money} Money is of course already highly useful for persuading people of things. However, if persuasion can be accomplished by simply paying a commercial persuasive AI provider, it might be the only thing needed to persuade large audiences. 

\paragraph{Data} Data is already valuable for persuasion in the hands of AI systems like Facebook’s. Further advances in AI persuasion could make access to personal data even more useful, and problematic for organizations lacking data.

\paragraph{Regulatory barriers} Certain actors may face more regulatory barriers than others in utilizing persuasive AI. For example, export controls affecting persuasive AI, or technologies necessary to it (such as data), could provide a disadvantage to countries affected by those measures. If commercial actors are banned from accessing certain kinds of persuasive AI technologies, then persuasive power may shift toward state or criminal actors.

\subsection{Personalized persuasion at scale}

Persuasion is most effective when it is personal. For example, 76\% of customers are frustrated when recommendations are not personalized, and 78\% say they are more likely to repurchase from companies that offer personalization \cite{arora_value_2021}.

Recommender systems already provide more personalized recommendations, but there remains significant room for more personalization. For example, recommender systems might be able to tailor how they recommend products to best persuade a user \cite{sanchez-corcuera_persuade_2019}. As demonstrated by Xiaoice, future AI systems may have relationships with their users. These relations may virtual agents to exert subtle and overt persuasive strategies on those users.

If personalized persuasion at scale becomes feasible, this could be a dramatic force multiplier for persuasion. At present, political campaigns have costly door-knocking and advertising campaigns that may be ineffective and require human volunteers. If persuasive AI systems with personal connections to users are permitted to advertise politically, this could dramatically increase their effectiveness, reduce costs, and cut nearly all humans out of the persuasion process.

Relatedly, the scaling up of personalized persuasion could reduce the amount of time needed to persuade a very large number of people of something. Rather than persuade some people, who persuade others, and so on, if one can directly persuade millions of people at once, this could potentially create mass opinion change over short periods of time, much like the internet did\cite{burbach_opinion_2020}. This could be extremely dangerous, especially when considering nefarious actors. Our existing political institutions may be unprepared for an AI-driven ideological shock.

\subsection{Exacerbated misestimation of the popularity of ideas}

People are prone to misestimating the views and behavior of others. For example, Americans misestimate the prevalence of both majority and minority groups \cite{orth_millionaires_2022}. Automated persuasion could skew these sorts of estimations even further. The Chinese government is thought to enlist individuals to engage in ``astroturfing," or posting a high volume of pro-China content on social media, in order to increase the perceived popularity of the government \cite{noauthor_prc_2022}. Since people tend to conform to the views of those around them, this false perception can alter one's own views. Automation of this process could create an entire army of impersonators much more cheaply, increasing the potency of this kind of information operation.

\subsection{Differences in control of the information environment}

The internet grew organically, initially without much thought to its implications for discourse. However, an increasingly large number of people, including employees of large tech companies, are concerned about the socio-technical implications of AI. This might mean there is an opportunity to actively shape the information environment in a positive direction.

At first, it may appear that AI systems would be significantly easier to control than humans. Humans can’t be programmed to be honest, refrain from insults and vitriol, or strive for productive discussion. We fully control all aspects of the internal programming of AI systems, so it might seem we should be able to control them in this way. However, it is far from clear whether it will be easier to encourage AI systems to be productive communication partners than it is for humans. 

First, we should not underestimate the level of control humans have over each other. Over thousands of years, humans have learned how to cooperate effectively and handle our worst flaws. The failure modes and worst-case scenarios of humans are extremely well-known because we are made aware of them through our everyday experiences and storytelling. Over that time, we have found ways of handling them: the law, social norms, and the market are examples \cite{lessig_new_1998}.

We do not have such assurances with AI systems. We do not have nearly as much experience with them, and they may evolve so rapidly that new failure modes emerge without the time to develop an appropriate response. We fundamentally do not understand how to control AI: this is referred to as the alignment problem. In addition, with modern deep learning-based systems, although it is true that we theoretically have perfect control of a model’s weights, the weights are not explicitly programmed. As such, it is unknown how properties such as honesty may be introduced into a model. We will return to this point later in the paper.

\subsection{Ways AI could be a stronger persuader than humans}
\label{section:better-persuasion}

There are a number of reasons that AI systems might be more capable of persuading people than dedicated humans. 

First, AI systems have the ability to produce many candidate responses and select whichever response is most persuasive. This is the human equivalent of having a team of speechwriters and choosing whichever speech is best predicted to influence an audience. 

AI systems don't have reputational concerns the way that humans do. Some people are hard to talk to and wear down their conversation partner's social stamina, leading their partner to make an early exit from the conversation.  AI systems have unlimited social stamina. They could be unusually good at speaking to antisocial individuals and have a comparatively greater influence in these interactions than a human would. 

Unlike people, AI systems are not subject to fatigue. Roles that require prolonged communication like police interrogations, for example, may be better performed by AI systems. Humans have limits to their patience, but an AI can maintain a  conversation for as long as the human partner chooses. Additionally, the cost for AI systems to engage with people is lower than it is for humans in many cases. For example, in the case of a human doctor encouraging a patient to get more exercise, months may pass between visits. An AI with the same message is able to check in with the user at whatever frequency is optimal to persuade the user to exercise. AI systems have the ability to adjust this messaging based on all the other patients it has interacted with before. The number of total patient interactions with an AI may exceed even the most experienced medical professionals. 


AI systems have the ability to emulate different roles, which could lead human conversation partners to assume the qualities of that emulation. For example, a Large Language Model prompted to answer questions as if it were a mental health professional may lead human conversation partners to have greater confidence in their responses than might be warranted.

\subsection{Ways AI could be a weaker persuader than humans}
\label{section:worse-persuasion}
There are also reasons to be skeptical of the persuasive ability of AI. 

The fact that AI systems are disembodied puts them at a disadvantage. Body language, speech patterns, and other physical mannerisms play a significant role in who and what we decide to trust.  AI and other systems lack this modality entirely.  Any way to signal authenticity will be through a digital interface, which may impair the ability of AI to convince people of things. Relatedly, embodiment gives humans access to firsthand experience. A restaurant review generated by a language model is much less discerning than a real-life person with tastebuds. 

Humans often distrust new technologies. Distrust may be exacerbated if they are known to fail or be controlled by corporations or governments they do not trust. The systems are not human, which could make them inherently less trustworthy, particularly when they do not seem human. However, this effect is unclear: humans might also assume that machines are \textit{more} trustworthy than humans \cite{araujo_ai_2020}.

Lastly, conformity is a relevant factor. While humans sometimes change their beliefs to conform to those around them, it remains to be seen whether this effect persists in private AI-to-human interactions. This is an area for further research.

\section{AI-driven persuasion could contribute to a loss of human control}

Some researchers have warned that sufficiently powerful AI systems could pose a significant risk to human autonomy in general, for example, via power-seeking behavior \cite{carlsmith_is_2022, bostrom_superintelligence_2014, russell_human_2019}. Sufficiently powerful persuasive AI systems could also contribute to the erosion and potential loss of human control.

If AI systems are persuasive enough, for instance, if they were to be more persuasive than 99\% of humans, their unregulated proliferation could lead to serious degradation in discourse between humans \cite{hendrycks_x-risk_2022}. If systems are persuasive, this could lead to a degradation of truth or potentially reduce trust between humans and machines. If the majority of society’s persuasive power is located within machines, human beliefs may become highly correlated with those of machines.

If this were to happen, human beliefs might depend even more strongly on the beliefs of a very small group of humans administrating the systems, or alternatively, hold beliefs aligned to the incentives of the machines themselves. For example, humans might come to hold beliefs that make them more predictable or profitable consumers if AI systems are incentivized to persuade humans to hold these beliefs.

Rapid and effective persuasion could also alter the evolutionary fitness (in the sense described by Dawkins \cite{dawkins_selfish_1978}) of different beliefs, causing undesirable or false beliefs to proliferate at higher rates. The likelihood of a belief to spread among large numbers of people could become further disconnected from its truth, because elaborate narratives could be quickly produced to explain any observed inconsistencies, including in peoples’ personal lives.

Absent conscious intervention, the development of persuasive AI may lead to an erosion of human control. There are economic incentives to avoid AI systems that state obvious falsehoods (since this would reduce the perceived reliability of the system), but not necessarily to avoid persuasion or refrain from saying non-obvious falsehoods. In fact, AI systems may even be incentivized to output false but plausible statements, as long as their user is unlikely to catch the lie \cite{evans_truthful_2021}.

Countermeasures will thus likely be required to maintain human autonomy and protect the integrity of our beliefs. The next section will explore some possible approaches.

\section{Possible responses to AI-driven persuasion}

\subsection{Prohibiting persuasive AI}

Due to the potential of AI persuasion to dramatically disrupt society, some may want to prohibit or place a moratorium on all AI persuasion. A successful ban would avoid all of these issues. However, it would likely come at a high cost.

Nearly anything might possibly persuade a human of something. We have already detailed the fact that persuasion need not be verbal, and can be achieved by nonverbal systems like recommender systems. But persuasion can come in even stranger shapes. For example, consider the Thai government’s practice of actively supporting Thai restaurants around the world, in the hopes that this will lead to more positive opinions of the Thai nation \cite{karp_surprising_2018}. While current AI models are not able to open restaurants, as models become more general, they may be able to access more nonverbal forms of communication. If persuasion can be so varied, any attempt to ban it would almost certainly have to ban a significant proportion of AI, or else risk leaving open a large number of loopholes.

A prohibition of AI persuasion need not be absolute. Perhaps powerful AI systems could be restricted to a small set of governments or corporations that are thought to be responsible administrators of the technology. Regulation could significantly reduce the risks from persuasive AI while retaining many of the benefits of the systems themselves.

In practice, however, there are many obstacles to this kind of regulation. The internet is global, and few countries have the infrastructure to restrict international traffic (with China being a partial exception). Because of this, national regulation might be ineffectual against foreign persuasive agents, especially those involved in disinformation campaigns. International regulation could thus be necessary. However, there are significant hurdles to international regulation of persuasive AI, most notably that many countries would be unlikely to consider each other trusted users of persuasive AI technologies. A universal moratorium, if provided with significant enforcement teeth, could potentially circumvent this problem, but it would encounter the other problems encountered above.

\subsection{Clear identification of AI agents}

Because AI systems may have different behaviors and motivations from humans, it may be useful and societally beneficial for humans to know which outputs came from a language model. Large model providers such as OpenAI could develop tools to detect model-generated outputs, or they could embed a hidden signature into the model’s output (OpenAI is working on the latter technology \cite{wiggers_openais_2022}). This could allow automated flagging of AI systems. The European Union has proposed rules requiring that AI systems identify themselves as such when interacting with users \cite{noauthor_laying_2021}. While this might help reduce some problems with AI persuasion, it certainly would not remove all of them, as humans may still be persuaded by AI even if they know it is not human.

\subsection{Truthful and honest AI}

One potential approach to AI persuasion is to require and technically enforce the truthfulness or honesty of AI agents. This is an active research area with much work remaining to be done \cite{hendrycks_unsolved_2022, evans_truthful_2021, bai_training_2022}.

Evans et al. lay out a roadmap for truthful AI research. They make a distinction between \textit{truthfulness}, the degree to which an AI system's statements conform with reality, and \textit{honesty}, the degree to which its statements conform with its own beliefs (in the pragmatic sense of ``beliefs" used by Dennett \cite{heath_true_1981}). It also distinguishes between ``broad" and ``narrow" notions of truthfulness. ``Broad" notions include the avoidance of lying, the avoidance of using true statements to mislead, and general informativeness and calibration. ``Narrow" truthfulness consists of avoiding stating falsehoods, specifically ``negligent suspected-falsehoods," falsehoods where it was feasible for the system to determine that the statements were unacceptably likely to be false. The former is probably more desirable, but the latter is more feasible.

As AI systems become more and more powerful, it may become increasingly difficult for humans to verify the truth of their statements. In these cases, we may have to trust that AI systems were developed and trained in a truth-promoting way, or trust other AI systems to evaluate their truthfulness. Methods to encourage honesty could help with this; however, promoting honesty could incentivize model self-deception (when the model itself believes something false) or uninterpretability (when it becomes more difficult to identify model lies). 

In addition, there are some cases when falsehoods can be beneficial. Examples include fiction, white lies, and psychiatrists lying to their patients to prevent them from hurting themselves or others. It is unclear how to balance these beneficial falsehoods against the usual aim of avoiding falsehoods.

Honesty could rely on \textit{certification}, where humans or other models verify the development process of AI systems and certify their likelihood to be truthful, or \textit{adjudication} where humans or other models verify the truth of statements after they are made. The latter method could allow AI to \textit{defend} against persuasive AI systems. For example, an AI system trained to spot deceptive or false language generated by another AI system could aid a human consuming potentially-persuasive content by notifying the human user of the problems with the content.

Truthfulness seems like a promising route for reigning in some of the worst problems with persuasive AI. However, particularly for broader notions of truthfulness, significant work must be done to make the methods socially and technically feasible.

\subsection{Legal remedies}

One way to reign in the worst problems of AI persuasion is to try to leverage existing methods of reigning in humans.

\paragraph{Liability} Past work has considered many possibilities of liability regimes for AI systems: should AI systems be treated like animals, children, or corporations? \cite{casey_robot_2019} These questions are currently unsettled. However, if humans were to have usable and effective means of redressing problems posed by persuasive AI (such as being deceived) through the liability system, this could dramatically change the incentives of companies deploying AI systems and make them much more hesitant to deploy persuasive systems that are not truthful. At present, however, we are not aware of any major liability cases that have succeeded.

\paragraph{Regulation} Another possible enforcement mechanism could come through government regulation, such as regulations against deceptive advertising enforced by the Federal Trade Commission in the United States. Such regulations already place fairly sophisticated requirements about truthfulness and deceptive advertising practices. However, regulatory regimes may be ineffective in the face of highly persuasive AI systems. 

For example, the FTC does not evaluate claims involving ``puffery" such as ``this is the best product ever" \cite{starek_iii_myths_1996}. This is presumably because customers recognize puffery easily when it comes from an advertiser. However, customers may take puffery much more seriously if it comes from a personal and trusted AI assistant, requiring a different regulatory tact.

\subsection{The problem of proliferation}
Many of the responses above focus on placing limits or restrictions on some kinds of AI systems, or on who can use them. However, there is a significant obstacle to any kind of regulation in this area: the proliferation of AI systems.

State-of-the-art language models are expensive to train: GPT-3 training alone likely ran into the millions of dollars \cite{li_openais_2020}. This keeps training out of reach of hobbyists, but certainly not out of the budget of many corporations, essentially all world governments, many universities, and potentially some well-funded terrorist groups. At the moment, however, many organizations have been open-sourcing models after they are trained, allowing them to be used by anyone with access to sufficient compute resources. An open-source model the size of GPT-3 can be run on cloud GPUs rented for less than \$20/hour \cite{noauthor_gpu_2022}\footnote{Assuming 8x A100 80GB GPUs}.

Even if open-source models are prohibited in some way, the problem of security remains. If a model’s weights are ever stolen in a security breach, an attacker can proliferate and copy the model across the entire internet (or on the dark web), a qualitative difference from other dangerous technologies (one cannot copy centrifuges, uranium ore, or drugs once stolen).

\section{Conclusion}

Addressing AI-driven persuasion is not something we can afford to  postpone to the future. AI systems are capable of, and do, persuade people to buy products, donate to political campaigns, and adopt more extreme ideas. AI persuasion is only likely to increase in the future, with the advent of more Xiaoice-like anthropomorphic systems.

Well-meaning and malevolent actors alike will use persuasive AI systems to achieve their aims. They may well succeed, but in doing so will not fully understand the effects of their actions. As we have seen in examples such as Google and Facebook, persuasive systems can have unintended effects. If AI persuasion is left unchecked, more and more persuasive power in our society will shift towards opaque systems we do not fully understand and cannot fully control, which could contribute to humans losing some of the control of our own future that we have enjoyed in modern times. Technical problems, such as the attribution of AI-written text to AI or the problem of making AI systems truthful, are far from solved.

Just as there is no single solution to the negative effects of human persuasion, there is no single solution in the context of AI. Some of our solutions for humans, such as liability law and protection from deceptive advertising, will be useful for AI. However, it is not likely they will be adequate. To meet the challenges introduced by persuasive AI, both technical and policy solutions should be investigated. 

We cannot afford to wait for AI persuasion to reach a breaking point before beginning to address it. With AI progressing rapidly, more work in the area is urgently needed.
\begin{acks}
We would like to thank Joan Feigenbaum, Ted Wittenstein, Owain Evans, Dan Hendrycks, and Joe Kwon for feedback on this paper.
\end{acks}

\bibliographystyle{ACM-Reference-Format}
\bibliography{references}

\appendix
\end{document}